\documentclass[aps,prc,preprint,showpacs,floatfix]{revtex4-1}  
\usepackage{graphicx}
\usepackage{amsmath}
\usepackage{color}

\newcommand {\nc} {\newcommand}
\nc {\beq} {\begin{eqnarray}} \nc {\eol} {\nonumber \\} \nc {\eeq}
{\end{eqnarray}} \nc {\eeqn} [1] {\label{#1} \end{eqnarray}} \nc
{\eoln} [1] {\label{#1} \\} \nc {\ve} [1] {\mbox{\boldmath $#1$}}
\nc {\rref} [1] {(\ref{#1})} \nc {\Eq} [1] {Eq.~(\ref{#1})} \nc
{\re} [1] {Ref.~\cite{#1}} \nc {\dem} {\mbox{$\frac{1}{2}$}} \nc
{\arrow} [2] {\mbox{$\mathop{\rightarrow}\limits_{#1 \rightarrow
#2}$}}
\begin{document}
\title{Peculiarity of the $^{12}$C$(0^+)$ and $^{12}$C$(2^+)$ energy  spectrum in a 3$\alpha$ model}
\author {E. M. Tursunov}
\email{tursune@inp.uz} \affiliation {Institute of Nuclear Physics,
Academy of Sciences, 100214, Ulugbek, Tashkent, Uzbekistan}
\author{I. Mazumdar}
\email{indra@tifr.res.in} \affiliation{$2$ Dept. of Nuclear $\&$
Atomic Physics, Tata Institute of Fundamental Research, Mumbai 400
005, India}
\begin{abstract}
 Lowest energy spectrum of the $^{12}$C nucleus is analyzed in the 3$\alpha$ cluster model with
 a deep $\alpha\alpha$-potential of  Buck, Friedrich and Wheatley with Pauli forbidden states in the $S$ and $D$
 waves. The direct orthogonalization method is applied for the elimination of the 3$\alpha$-Pauli forbidden states.
 The effects of possible first order quantum phase transition are shown in the lowest $^{12}$C($0_1^+)$ and  $^{12}$C($2_1^+)$
 states from  weakly bound phase to a deep phase. The ground and lowest
 $2^+$ states of the $^{12}$C nucleus in the deep phase are created by the critical eigen states of the
 Pauli projector for the $0^+$ and $2^+$ three-alpha functional spaces, respectively.
\end{abstract}

 \maketitle

\section{Introduction}

\par The studies in structure and structural evolution of atomic nuclei
continue to be one of the major aspects of low and medium energy
nuclear physics.  The richness of nuclear structure and the spectrum
of various modes of excitations have led to the development of
different models to understand the underlying mechanisms guiding the
response of the nucleus. The plethora of nuclear models developed
from as early as the thirties have been guided by experimental data
that show both single particle and collective behavior of the
nuclei. In addition, the richness is further enhanced by the
observation of interplay of single particle and collective behavior.
The evolution of nuclear structure with physical observables like,
angular momentum, isospin, temperature are associated with phase
transitions, both continuous and discontinuous. The studies in
different types of phase transitions happen to be one of the central
themes of nuclear structure studies. These include both Quantum
Phase Transitions (QPT) at zero temperature
\cite{serdar16,pavel2014,pavel2015,pavel2019} and phase transitions
associated with nuclei at finite temperature and angular momentum
\cite{goodman94,goodman01,mazumdar09,mazumdar17}. Abrupt changes in
the ground state shapes from spherical or very small deformations to
large deformed shapes with change in neutron (N) numbers for a given
element have been observed and theoretically explained to be due to
quantum phase transitons \cite{toga16}. Such changes are associated
with massive reorganization of the proton and neutron orbitals and
is understood in terms of a QPT.  Variation in some non-thermal
control parameter (say, number of nucleons) is responsible for
inducing such abrupt phase transitions. QPT driven by variation in
some defined order parameter has also been studied in cases of
spontaneous fission of heavy nuclei \cite{malik19}.
%
%

The behavior of atomic nuclei have resulted in variety of nuclear
models. The alpha cluster model has been associated with the light
even-even nuclei which can justifiably be modeled as an ensemble of
alpha particles. The $^{12}$C nucleus occupies a pre-eminent
position in the list of nuclei which support alpha-cluster states.
Arguably, the 7.65 MeV, 0$^{+}$ excited state, well known as the
Hoyle state, is the most famous cluster state of the atomic nucleus.
There has been a large body of work trying to describe the cluster
states in nuclei like $^{12}$C and other self conjugate nuclei like
$^{16}$O, $^{20}$Ne etc. One of the interesting properties of these
nuclei is their special structure, associated with the Bose-Einstein
condensation \cite{review12C}. Another special structure is
connected with the QPT found in the {\it ab-initio} calculations
\cite{serdar16}. Broadly, the cluster models can be categorized into
microscopic and macroscopic approaches with their fair share of
success, simplicity, failure and difficulties.  On the one side,
within the microscopic models the ground and the first 2$^+$ excited
states of $^{12}$C are strongly overbound by about 4-6 MeV in
comparison with the experimental energy values \cite{vas12}. Only a
complicated nonlocal $\alpha \alpha$ potential derived from the
resonating group model calculations is able to reproduce the
energies of the ground state and the Hoyle ($0^+_2)$ resonance
\cite{suz08}. The macroscopic models treat the system like $^{12}$C
as an ensemble of structure-less alpha particles and use both
shallow and deep local potentials and also non-local potentials for
the binary $\alpha-\alpha$ systems.  The success of these methods in
producing the $\alpha-\alpha$ phase shifts and $^8$Be resonances and
difficulties in terms of producing experimental binding energies and
removing spurious, redundant states  in the $^{12}$C have been
studied and discussed in detail by several authors
\cite{rev,review12C,tur03,fuji06,suz08}.
%
%

Although the 3$\alpha$ cluster model for the structure of the lowest
$^{12}$C states seems very natural due to strong binding of nucleons
inside the $\alpha$-clusters, there are serious problems, associated
with a realistic modeling of Pauli forces. Repulsive local $\alpha
\alpha$-potentials, both l-dependent and l-independent, strongly
underestimate the bound states of the $^{12}$C nucleus \cite{tur03}.
The application of the alternative local deep $\alpha\alpha$-
interaction potential of Buck-Friedrich-Wheatley (BFW) \cite{BFW}
requires a careful treatment of the Pauli forbidden states (FS). The
method of orthogonalizing pseudopotentials (OPP) \cite{kuk78} is a
powerful technic for the elimination of forbidden states in a
three-body system. The wave functions of the $^6$He and $^6$Li
nuclei calculated in the $\alpha+N+N$ three-body model based on the
OPP method, have been successfully applied to the study of the beta
decay of $^6$He halo nucleus into the $\alpha +d$ continuum
\cite{tur06,tur06a}, and the astrophysical capture reaction
$\alpha+d \to ^6$Li $+\gamma$ \cite{tur16,bt18,tur18}. However, the
3$\alpha$ quantum system is strongly different from the
$\alpha+N+N$, which contains a single $\alpha$-particle. Here
$\alpha \alpha$-Pauli forbidden states play a decisive role in the
description of dynamics of the  3$\alpha$ system. Indeed, within the
OPP method it was found \cite{tur01,fuji06} that the energy spectrum
of the ground $0_1^+$ and first excited $2^+_1$ states is highly
sensitive to the description of the $\alpha \alpha$-Pauli forbidden
states. From the results of the calculations it was not possible to
understand how to fix the energies of the ground and excited levels,
since a convergence in respect to the projecting constant $\lambda$
was not clear. When passing values of $\lambda=10^4 - 10^6$ MeV the
energies of the $0^+$ states show a non-analytical behaviuor: for
the $\lambda=10^4$ MeV the energies of lowest states -16.106 MeV,
-0.422 MeV, 1.353 MeV change to the values -0.435 MeV, 1.407 MeV and
3.316 MeV for $\lambda=10^6$ MeV. In other words, the lowest state
with the energy -16.106 MeV was lost, which is not usual. The
energies of the lowest $2^+$ states show similar behaviour. Again
the lowest state with the energy $E(2^+_1)=-15.649$ was lost.

For understanding the enigmatic behaviour of the $^{12}$C(0$^+$)
spectrum the authors of Ref. \cite{fuji06} have applied more
transparent direct orthogonalization method of eliminating Pauli
forbidden states from the 3$\alpha$ functional space. They have
found that there are two so-called almost forbidden states (AFS) as
eigen states of the three-body Pauli projector, which play a
decisive role for the $^{12}$C(0$^+$) spectrum. So, if these AFS are
included into the 3$\alpha$ allowed functional subspace, then the
energy of the ground state is about -20 MeV, while it is $-0.20$ MeV
if these AFS are accepted as forbidden 3$\alpha$ basis states. In
order to avoid the AFS problem, the authors of above work suggested
to use the $\alpha \alpha$- forbidden states derived from underlying
microscopic theory and not to use the FS of the BFW potential. Such
a way gives normal three-body FS (as in other three-body systems
like $\alpha+2N$) contrary to the three-body FS derived from the
initial $\alpha \alpha$-potential. However, they still yield a
strong overbinding of the 3$\alpha$ ground state. Moreover, from
physical viewpoint, this way is not realistic, since the forbidden
states should be associated with two-body potentials which describe
the experimental data, energy spectrum of bound and resonance
states, and phase shifts. Since the BFW potential yields a very nice
description of the experimental data for the $\alpha \alpha$-
scattering and the $^8$Be resonances, the specific properties of the
3$\alpha$ spectrum associated with the Pauli projecting could be
connected with a strong physics which is still not well understood.
%

The aim of present work is
to study peculiar properties of the $^{12}$C$(0^+)$ and
$^{12}$C$(2^+)$ energy spectrum associated with removing Pauli
forbidden states from the 3$\alpha$ functional space. A deep $\alpha
\alpha$-potential of BFW will be employed. Differently from the
Faddeev equation method in Ref. \cite{fuji06} we use a variational
method on symmetrized Gaussian basis. For the elimination of the
3$\alpha$ Pauli forbidden states we use the same direct
orthogonalization method from Ref. \cite{fuji06} where only
$^{12}$C(0$^+$) lowest states have been studied. We will examine a
similarity of the  $0^+$ and $2^+$ spectrum including the Hoyle
band. As a possible origin of above mentioned non-analytical
behaviour of the $^{12}$C spectrum, consequences of the QPT  in the
$^{12}$C nucleus will be discussed.

The theoretical model is described in Section 2. Sections 3 and 4
contain the numerical results for the $^{12}$C(0$^+$) and
$^{12}$C(2$^+$) spectrum, respectively. A discussion of the results
is given in Section 5 and conclusions are drawn in the last section.

\section{Theoretical model}

The direct orthogonalization method \cite{fuji06} is based on the
separation of the complete Hilbert functional space into two parts.
The first subspace $L_Q$, which we call allowed subspace, is defined
by the kernel of the complete three-body projector. The rest
subspace $L_P$ contains 3$\alpha$ states forbidden by the Pauli
principle. After the separation of the complete Hilbert functional
space of 3$\alpha$ states into the $L_Q$ (allowed) and
$L_P$(forbidden) subspaces, at next step we solve the three-body
Schr\"{o}dinger equation in $L_Q$.

The $\alpha\alpha$- interaction potential of Buck-Friedrich-Wheatley
(BFW) \cite{BFW} has a simple Gaussian form
\begin{equation}
V(r)=V_0 exp(-\eta r^2)+4e^2erf(br)/r,
\end{equation}
with parameters $V_0$=-122.6225 MeV, $\eta=0.22$ fm$^{-2}$ for the
nuclear part and  $b$=0.75 fm$^{-1}$ for the Coulomb part. This
choice of the potential parameters yields a very good description of
the experimental phase shifts $\delta_L(E)$ for the $\alpha \alpha$-
elastic scattering in the partial waves $L=0,2,4$ within the energy
range up to 40 MeV and the energy positions and widths of the
$^8$Be resonances.
%
Hereafter we use a value $\hbar^2/m_{\alpha}=10.4465$ MeV fm$^2$ for
comparison with the results of Ref.\cite{fuji06}. This potential
contains two Pauli forbidden states in the $S$ wave with energies
$E_1=-72.6257$ MeV and $E_2=-25.6186$ MeV, and a single forbidden
state in the $D$ wave with $E_3=-22.0005$ MeV. For the realistic
description of the system one has to eliminate all FS from the
solution of the three-body Schr\"{o}dinger equation.

\par The three-body Hamiltonian in the 3$\alpha$-cluster model reads:
 \begin{equation}
 \hat{H}=\hat{H}_0
+V(r_{23})+V(r_{31})+V(r_{12}),
\end{equation}
where $\hat{H}_0$ is the kinetic energy operator and $V(r_{ij})$ is
the interaction potential between the  $i$-th and $j$-th particles.
A solution of the Schr$\ddot o$dinger equation
\begin{equation}
 \hat{H}\Psi_s^{JM}=E\Psi_s^{JM}, \,\, \Psi_s^{JM} \in L_Q.
\end{equation}
should belong to the allowed subspace $L_Q$ of the complete
3$\alpha$ functional space.

The wave function of the 3$\alpha$- system is expanded in the series
of symmetrized Gaussian functions \cite{tur01}:
\begin{equation}
\Psi_s^{JM}=\sum_{\gamma j} c_j^{(\lambda ,l)}\varphi_{\gamma j}^s ,
\end{equation}
where $\varphi_{\gamma j}^s=\varphi_{\gamma
j}(1;2,3)+\varphi_{\gamma j}(2;3,1)+ \varphi_{\gamma j}(3;1,2) ,$
\begin{equation} \varphi_{\gamma
j}(k;l,m)=N_j^{(\lambda l)}x_k^{\lambda}y_k^lexp(-\alpha_{\lambda j}
x_k^2-\beta_{l j}y_k^2){\cal F}_{\lambda l }^{JM}
(\widehat{\vec{x}_k},\widehat{\vec{y}_k})
\end{equation}
Here $(k;l,m)={(1;2,3),(2;3,1),(3;1,2)}$, $\gamma =(\lambda
,l,J,M)=(\gamma_0,J,M); \vec{x}_k, \vec{y}_k $ are the normalized
Jacobi coordinates in the $k$-set:  $$
\vec{x}_k=\frac{\sqrt{\mu}}{\hbar} (\vec{r}_l-\vec{r}_m)\equiv
\tau^{-1}\vec{r}_{l,m} ; $$ \begin{equation}
\vec{y}_k=\frac{2\sqrt{\mu}}{\sqrt{3}\hbar} (\frac{\vec{r}_l+
\vec{r}_m}{2}-\vec{r}_k)\equiv \tau_1^{-1}\vec{\rho}_k ,
\end{equation}
$N_j^{(\lambda l)}$ is a normalizing multiplier. The nonlinear
variational parameters
 $\alpha_{\lambda j}, \beta_{l j}$ are chosen as the nodes of the
Chebyshev grid:
$$ \alpha_{\lambda j}=\alpha_0tg(\frac{2j-1}{2N_{\lambda}}\frac{\pi}{2}),
j=1,2,...N_{\lambda}, $$
\begin{equation}
\beta_{l j}=\beta_0tg(\frac{2j-1}{2N_{l}}\frac{\pi}{2}),
j=1,2,...N_{l},
\end{equation}
where $\alpha_0$ and $\beta_0$ are scale parameters for each
$(\lambda l)$ partial component of the complete wave function.
%
%
The angular part of the Gaussian basis is factorized as:
\begin{equation}
{\cal F}_{\lambda l}^{JM}(\widehat{\vec{x}_k},\widehat{\vec{y}_k})=
\{Y_{\lambda}(\widehat{\vec{x}_k}) \bigotimes
Y_l(\widehat{\vec{y}_k})\}_{JM}
 \phi(1) \phi(2) \phi(3),
\end{equation}
where $\phi(i) $ is the internal wave functions of the
$\alpha$-particles. Here the orbital momenta $\lambda$ and $l$ are
conjugate to the Jacobi coordinates $\vec{x}_k$ and $\vec{y}_k$,
respectively.

\par The kinetic energy operator of the Hamiltonian can be expressed
in the normalized Jacobi coordinates in a simple form as
\begin{equation}
\hat{H}_0=-\frac{\partial^2} {\partial\vec{x}_k^2}
-\frac{\partial^2} {\partial\vec{y}_k^2}
\end{equation}
within any choice of $(\vec{x}_k,\vec{y}_k)$, $k=1,2,3$. The matrix
elements of the kinetic energy operator and the interaction
potentials have been given in Ref. \cite{tur94}.
%
%

\par In order to separate the complete 3$\alpha$ functional space into
the $L_Q$ and $L_P$ we calculate eigen states and corresponding
eigen values of the projecting operator \cite{fuji06}
\begin{equation}
 \hat{P} =\sum_{i=1}^3\hat{P}_i,
\end{equation}
where each $\hat{P}_i, (i=1,2,3)$  is the sum of Pauli projectors
$\hat{\Gamma}_i^{(f)}$ on the partial $f$ wave forbidden states
($1S$, $2S$, and $1D$) in the i-th $\alpha \alpha$-subsystem:
\begin{equation}
\hat{P}_i= \sum_{f}\hat{\Gamma}_i^{(f)},
\end{equation}
\begin{equation}
\hat{\Gamma}_i^{(f)}=\frac{1}{2f+1} \sum_{m_f} \mid \varphi_{f
m_f}(\vec{x}_i)> < \varphi_{f m_f}(\vec{x'}_i) \mid
\delta(\vec{y}_i-\vec{y'}_i) ,
\end{equation}
 with the
forbidden state function expanded into the Gaussian basis:
\begin{equation}
 \varphi_{f m_f}(\vec{x}_i) =x_i^f \sum_m
N_m^{(f)}b_m^{(f)}exp(-\frac{r_i^2}{2r_{0m}^{(f)2}})
Y_{fm_f}(\hat{\vec{x}}_i).
\end{equation}
Here $r_0$ is the "projector radius" and $N_m^{(f)}$ is the
normalizing multiplier:
\begin{equation}
 N_m^{(f)}=2^{f+7/4}
\frac{\alpha_m^{(2f+3)/4}}{\pi ^{1/4}[(2\lambda+1)!!]^{1/2}}, \qquad
\alpha_m=\tau^2/(2 r_{0m}^2).
\end{equation}

\section{$^{12}$C(0$^+$) spectrum}

First we calculate the $^{12}$C$(0^+)$ spectrum. The three body
channels $(\lambda,\ell)=(0,0),(2,2),(4,4)$ contain up to 280
Gaussian functions. Convergence is fast due to the use of
symmetrized basis functions. We reproduced the spectrum of the
operator $\hat{P}$ in the 3$\alpha$ functional space with
$J^{\pi}=0^+$, firstly calculated in Ref.\cite{fuji06}. It belongs
to the interval from 0 to 3. As was noted above, the allowed
subspace $L_Q$ is defined by the eigen states of the operator $\hat
P$, corresponding to its zero eigen value: $L_Q=kern(\hat P)$.
However, in the 3$\alpha$ system this procedure is not easy due to a
high sensitivity of the energy on the description of  $\alpha
\alpha$-Pauli forbidden states \cite{tur01}. As in mentioned work,
there are two eigen states of the operator $\hat{P}$ among other
eigen states, which play a decisive role for the structure of the
$^{12}$C(0$^+$) lowest states. The first special eigen state
$\Phi_1$ corresponds to a small eigen value
$\epsilon_1$=1.35333$\times 10^{-5}$: $\hat{P} \Phi_1 = \epsilon_1
\Phi_1 $.
\begin{figure}[htb]
\includegraphics[width=100mm]{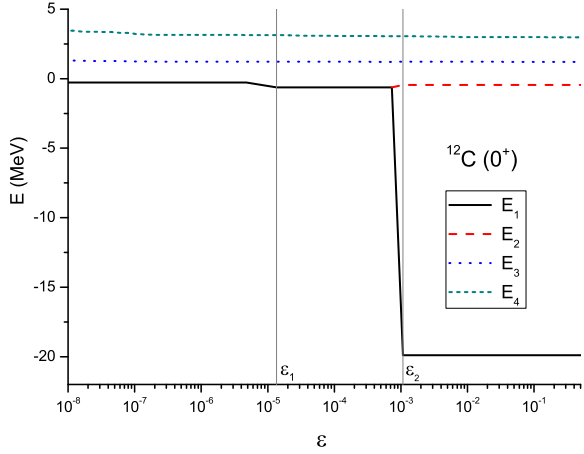}
\caption{Energy spectrum of the lowest $^{12}$C($0^+$) states in
dependence on the maximally allowed eigen values $\epsilon$ of the
operator $\hat P$.} \label{0+}
\end{figure}
The second eigen state $\Phi_2$ of the operator $\hat{P}$ with
corresponding eigen value $\epsilon_2$=1.07152$\times 10^{-3}$ is
especially important for the structure of the $^{12}$C nucleus. Both
eigen values are close to, but not exactly zero, nevertheless they
strongly influence on the structure of the carbon nucleus. This
makes the corresponding eigen states of the operator $\hat{P}$ very
special and in above work \cite{fuji06} they are called "almost
forbidden states" (AFS). In Fig.\ref{0+} we display the calculated
0$^+$ spectrum of the lowest $^{12}$C states as a function of
$\epsilon$, the maximal allowed eigen value of the operator
$\hat{P}$. As can be seen from the figure, the first AFS $\Phi_1$
influences only the lowest $^{12}$C(0$^+$) spectrum. Other levels
are not affected by the three-body projector, indicating that they
belong to the continuum spectra or correspond to a resonance
\cite{fun06}. The first AFS $\Phi_1$ decreases the lowest 0$^+$
energy from -0.278 MeV to -0.627 MeV. The next AFS $\Phi_2$ creates
a new 3$\alpha$ state with the energy of -19.897 MeV, which becomes
the ground state of the $^{12}$C nucleus. It is strongly bound with
binding energy of about 20 MeV or underbound with the energy
$E=-0.627$ MeV in respect to the 3$\alpha$ threshold in dependence
of that $\Phi_2$ belongs to $L_Q$ (allowed) or $L_P$ (forbidden). In
Ref. \cite{fuji06} the situation was not accepted as physically
possible. On the other hand, a non-analytical behavior of the lowest
$^{12}$C($0^+$) states around the critical point $\Phi_2$ can be
interpreted as a first-order quantum phase transition (QPT) from the
weakly bound phase  $\Psi_0$ with the energy -0.627 MeV to the deep
phase $\Psi_1$ with the energy of -19.897 MeV. The situation is very
close to the finding of Ref. \cite{serdar16} stating about the
"nature near a quantum phase transition". Indeed, the QPT occurs
due-to the quantum fluctuation between the two phases, which are
rooted in the Heisenberg uncertainty principle \cite{bookQPT}.

Beyond the critical point $\Phi_2$ the energy of the Hoyle state
$\Psi_2$ increases to $E=-0.458$ MeV, which is, however, lower than
the experimental energy value $E_{exp}(0_2^+)$=0.380 MeV. Also it
can be found that on the left side of the critical point $\Phi_2$ in
Fig. \ref{0+} the ground and Hoyle states coexist in the preliminary
weakly bound phase  $\Psi_0$. The overlap of $\Psi_0$ with the Hoyle
state $\Psi_2$ is 0.992, while its overlap with the ground state in
the deep phase  $\Psi_1$ is 0.109. This means that the ground state
deep phase is mostly created by the critical $\Phi_2$ eigen state of
the operator $\hat P$ from the continuum. The theoretical energy of
-19.897 MeV is significantly lower than the experimental value
$E_{exp}=$-7.274 MeV \cite{energylevel}, which indicates that the
ground state can be in the deep phase $\Psi_2$ with a probability,
smaller than 1. The energy of the Hoyle state can be matched to its
experimental value with the help of a weakly repulsive three-body
potential. At the same time, the $0^+_3$ resonance energy at
$E^*=10.3$ MeV \cite{energylevel} is well reproduced without any
additional potential. In Fig. 1 it corresponds to the state with the
energy $E_4=3.058$ MeV.

\section{$^{12}$C(2$^+$) spectrum}

\begin{figure}[htb]
\includegraphics[width=100mm]{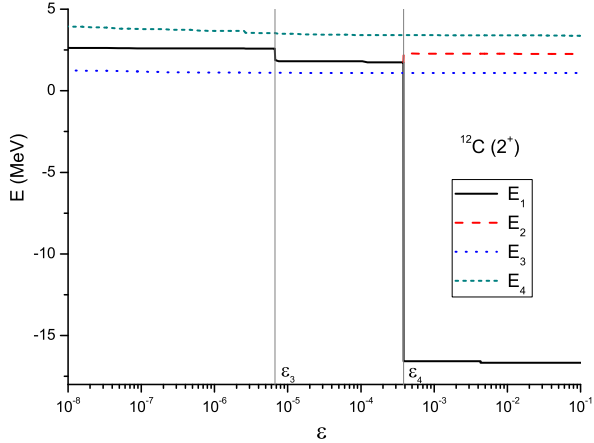}
\caption{Energy spectrum of the lowest $^{12}$C($2^+$) states in
dependence on the maximally allowed eigen values $\epsilon$ of the
operator $\hat P$.} \label{2+}
\end{figure}

Now we go to the $^{12}$C$(2^+)$ spectrum. In this case the $2^+$
functional space of the 3$\alpha$ system is built in the three-body
channels $(\lambda,\ell)=(0,2),(2,0),(2,2)$. Numerical calculations
have been done with up to 340 symmetrized Gaussian basis functions
which yield a good convergence of the results. Exactly as in the
case of the $0^+$ spectrum, the projector $\hat P$ contains two AFS
$\Phi_3$ and $\Phi_4$ with corresponding eigen values
$\epsilon_3$=6.74419$\times 10^{-6}$ and $\epsilon_4$=3.83029$\times
10^{-4}$, which play a decisive role for the $^{12}$C$(2^+)$
spectrum. As can be seen in Fig. \ref{2+}, a behavior of the $2^+$
spectrum is very close to the behavior of the lowest $0^+$ states.
The eigen state $\Phi_3$ of the operator $\hat P$ corresponding to
$\epsilon_3$, changes the second $2_2^+$ state energy from 2.578 MeV
to 1.873 MeV. Again as in the previous case, the critical point
$\epsilon_4$ (eigen state $\Phi_4$ of the operator $\hat P$) creates
a new deep phase with the energy $E=$-16.572 MeV, much lower than
the experimental energy value $E_{exp}(2_1^+)$=-2.834 MeV. In other
words, we again have a possible first order quantum phase transition
from weakly bound phase  $\Psi_0(2^+)$ to the deep phase
$\Psi_1(2^+)$. On the right-hand side of the critical point $\Phi_4$
(or $\epsilon_4$= 3.83029$\times 10^{-4}$) the Hoyle analog state
$\Psi_2(2^+)$ energy increases to $E=$2.279 MeV, slightly lower than
the experimental value $E_{exp}(2_2^+)$=2.596 MeV. Again as in the
case of the $0^+$ spectrum, on the left side of the critical point
$\Phi_4$ in Fig. \ref{2+} the lowest $2_1^+$ and Hoyle analog
$2_2^+$ states coexist in the preliminary weakly bound phase
$\Psi_0(2^+)$. The overlap of $\Psi_0(2^+)$ with the Hoyle analog
state $\Psi_2(2+)$ is 0.876, while its overlap with the lowest
$2_1^+$ phase $\Psi_1(2^+)$ is -0.334. The two $2^+$ levels in Fig.
\ref{2+} with energies $E_3$ and $E_4$ are not affected by the
projecting procedures, hence belong to the 3$\alpha$ continuum
spectrum.

\section{Discussion}

The energy of the Hoyle state $E(0^+_2)=0.38$ MeV \cite{energylevel}
can be reproduced with the help of a weak three-body potential
$V(\rho)=23 \, exp(-0.1 \rho^2)$ MeV, where $\bf{\rho}$ is the
hyperradius. The same potential yields for the Hoyle analog state
energy a value $E_{th}(2_2^+)=2.567$ MeV which is very close to its
experimental value of 2.596 MeV.

 Now we can discuss the situation
that happened in Ref. \cite{tur01} when applying the OPP method for
the elimination of Pauli forbidden states in the 3$\alpha$ system.
In this method the Hamiltonian reads
\begin{equation}
 \tilde{H}=H_0
+\tilde{V}(r_{12})+\tilde{V}(r_{23})+\tilde{V}(r_{31}),
\end{equation}
where the pseudopotentials
\begin{equation}
\widetilde{V}(r_{ij})=V(r_{ij}) +\sum_f\lambda_f
\hat{\Gamma}_{ij}^{(f)}.
\end{equation}
Here $\lambda_f$ is the projecting constant,
$\hat{\Gamma}_{ij}^{(f)}$ is the projecting operator to the $f$-wave
forbidden state in the two-body subsystem $(i+j)$,
$(i,j,k)=(1,2,3)$, and their cyclic permutations.

When increasing $\lambda_f$ one should have more and more repulsive
interaction and all the forbidden states should go out of the
allowed three-body functional space. The sensitivity of the energy
on alpha-alpha forbidden states can be seen from Tables 2 and 3 of
Ref.\cite{tur01} for the values of projecting constant
$\lambda_f=10^4-10^6$ MeV. The method of OPP allows to eliminate all
the Pauli forbidden states in the 3-body system with the increasing
projecting constant $\lambda_f$ up to infinity. This works well for
the $\alpha+N+N$ system, but does not work for the 3$\alpha$ system.
Indeed, when increasing projecting constant $\lambda_f$ from 10$^4$
MeV to 10$^6$ MeV, the energy of the ground state changes from
-16.106 MeV to -0.435 MeV, and the energy of the 2$_1^+$ state
changes from -15.65 MeV to a positive energy value (the binding was
lost).

In previous sections of present paper, as well as in
Ref.\cite{fuji06} for understanding the strong sensitivity of the
3$\alpha$ bound state energies on the description of the Pauli
forbidden states a different, direct orthogonalization method was
applied which is based on the separation of the complete Hilbert
functional space into two parts: allowed and forbidden subspaces.
This  method is more transparent than the OPP method. Here we can
understand well what is happening in the OPP method when increasing
$\lambda_f$ to infinity.  The situation was understood as coming
from the so called "almost forbidden states" (AFS) which play a
decisive role for the spectrum of the  12C nucleus. When $\lambda_f$
goes to infinity, one completely removes these "almost forbidden
states" from the functional model space of relative motion. The
situation is in strong contrast to other systems like $^6 \rm
Li=\alpha+p+n$, $^6\rm He=alpha+n+n$, where one has a convergent
energy value with increasing $\lambda_f$ to infinity. In other
words, in the 3$\alpha$ system, if we remove these AFS from the
model space, then we have a strong underbinding. And contrary, if we
include these AFS into the model space, then we have a strong
overbinding.

Thus we can state that the above effect, which we treat as possible
quantum phase transition, is not an artifact of the specific
procedure used. Since the effect was seen also with the OPP method
earlier, but was fully understood only with a more transparent
direct orthogonalization method from Ref. \cite{fuji06}. And the
Hamiltonian of OPP method contains a parameter $\lambda_f$
(projecting constant): when increasing it one has more repulsive
interaction, when decreasing it one has more attractive interaction.
In the direct orthogonalization method  we have a parameter
$\epsilon$, the maximal allowed eigen value of the operator $\hat
P$: when increasing it one has more attractive interaction, when
decreasing it one has more repulsive interaction in the Hamiltonian.
In this sense we state that the situation is very close to the
finding of Ref. \cite{serdar16}

An important and difficult question is, how the above possible
quantum phase transition effects in the $^{12}$C($0_1^+)$ and
$^{12}$C($2_1^+)$ lowest states can be directly detected in the
experiment. For the nuclear interaction time of order 10$^{-23}$ -
10$^{-22}$ s the quantum fluctuation can occur with the energy shift
$\Delta E \approx \hbar/\Delta t \approx$ 6.6 - 66 MeV. Our model
calculations gives an energy shift values of about 19.27 MeV and
18.45 MeV for the possible QPT in the $^{12}\rm C(0^+)$ and $^{12}
\rm C (2^+)$ spectrum, respectively. These estimations are very
consistent with above (6.6 - 66) MeV interval for the energy shift
of quantum fluctuations.

\section{Conclusion}

In summary, the energy spectrum of the $^{12}$C nucleus has been
analyzed within the 3$\alpha$ model. The Pauli forbidden states were
treated by the exact orthogonalization method. An evidence of
possible first order quantum phase transition has been examined. It
was shown that there are effects of possible QPT in the lowest
$^{12}$C($0_1^+)$ and $^{12}$C($2_1^+)$ states from the weakly bound
phase to a deep phase. For the $0^+$ spectrum there is a critical
eigen function (critical point) and corresponding critical eigen
value of the three-body projector, which is responsible for the
quantum phase transition. On the left hand side of the critical
point the lowest $0^+$ state mostly presents the Hoyle state, while
on the right hand side of the critical point the lowest state
becomes the ground state of the $^{12}$C nucleus in the deep phase.
An overlap of the critical eigen function of the three-body
projector with the ground state is close to unity, while its overlap
with the Hoyle state is almost zero. This means that the ground
state of the $^{12}$C nucleus in the deep phase  is created by the
critical eigen function of the Pauli projector. A behavior of the
$2^+$ levels is analogous. The lowest $2^+$ state in a deep phase is
created by the critical eigen function of the Pauli projector for
the $2^+$ levels.

Main physical result is that the origin of a possible quantum phase
transition is strong Pauli forces in the  3$\alpha$ system. If one
calculates the spectrum of the 3$\alpha$, 4$\alpha$, 5$\alpha$ , etc
quantum systems in the alpha-cluster model, the same  QPT effects
can be seen.  These QPT effects are not specific due-to the
orthogonalization method, but rather due to the Pauli forces. They
can be seen in any alpha-cluster model with exact treatment of the
Pauli principle.

\end{document}